\documentclass{PoS}

\usepackage{amsmath}
\usepackage{amssymb}
\usepackage{comment}
\usepackage{epsfig,latexsym,amsfonts,amsmath,amsthm,amssymb,amsbsy,multirow,slashed,color,mathrsfs,wasysym,textcomp,subfigure,wrapfig,comment,bbold,array,longtable,multirow,bbm}

\usepackage{tikz}
\usetikzlibrary{positioning}
\usetikzlibrary{intersections} 

\newlength{\PicScale}
\newcommand{\ZZ}{{\mathbb{Z}}}
\newcommand{\PP}{{\mathbb{P}}}

\newcommand{\kod}[1]{\mathrm{#1}}
\newcolumntype{M}[1]{>{\centering\arraybackslash}m{#1}}
\newcolumntype{N}{@{}m{0pt}@{}}

\title{Non-geometric heterotic backgrounds and 6D SCFTs/LSTs}

\ShortTitle{Non-geometric heterotic backgrounds and 6D SCFTs/LSTs}

\author{Anamar\'ia Font\\
       Facultad de Ciencias, Universidad Central de Venezuela, A.P.20513, Caracas 1020-A, Venezuela\\
       E-mail: \email{afont@fisica.ciens.ucv.ve}}

\author{I\~naki Garc\'ia-Etxebarria\\
       Max-Planck-Institut f\"ur Physik, F\"ohringer Ring 6, 80805  M\"unchen, Germany\\
       E-mail: \email{inaki@mpp.mpg.de}}

\author{Dieter L\"ust\\
       Max-Planck-Institut f\"ur Physik, F\"ohringer Ring 6, 80805 M\"unchen, Germany; and\\
        ASC for Theoretical Physics,
        Theresienstra\ss e 37, 80333 M\"unchen, Germany\\
       E-mail: \email{dieter.luest@lmu.de}}        

\author{Stefano Massai\\
       Enrico Fermi Institute, University of Chicago, 5640 S Ellis Ave, Chicago, IL 60637, USA\\
       E-mail: \email{massai@uchicago.edu}}
       
\author{\speaker{Christoph Mayrhofer}
        \\
        ASC for Theoretical Physics,
        Theresienstra\ss e 37, 80333 M\"unchen, Germany\\
        E-mail: \email{christoph.mayrhofer@lmu.de}}

\abstract{%
We study ${\mathcal N}=(1,0)$ six-dimensional theories living on defects of
non-geometric backgrounds of the $E_8\times E_8$ 
and the $\text{Spin}(32)/{\mathbb Z}_2$ heterotic strings. 
Such configurations can be analyzed by dualizing to F-theory 
on elliptic K3-fibered non-compact Calabi-Yau threefolds. 
The majority of the resulting dual threefolds turn out to contain singularities which do not admit a crepant resolution. When the singularities can be resolved crepantly,  
the theories living on the defect are explicitly determined and reveal a 
form of duality in which distinct defects are described by the same 
IR fixed point. In particular, a subclass of non-geometric defects 
corresponds to SCFTs/LSTs arising from small heterotic instantons on ADE
singularities. %
}

\FullConference{Corfu Summer Institute 2016 "School and Workshops on Elementary Particle Physics and Gravity"\\
		31 August - 23 September, 2016\\
		Corfu, Greece}

\begin{document}

\section{Introduction}
Since the early studies of string compactifications most work has been done in the supergravity regime. 
However it is well known that string vacua can be much richer.  
Our motivation in this article is to take a step away from vacua  where the
background can be understood geometrically by considering a classical
supergravity compactification. The aim is both to learn more about the
non-classical and non-geometrical properties of string theory, and to gain some insight
about the broader set of allowed string vacua.

In this paper we consider a class of heterotic string vacua which are very non-classical,
involving compactifications on ``spaces'' that cannot be globally
described as geometries, while remaining accessible thanks to duality
with F-theory  \cite{Vafa:1996xn}. In this way we can probe many of the properties of the
heterotic string away from the classical regime where it is usually studied.
More concretely, we will focus on cases where the compactification
space for the heterotic string at a generic point  is locally
geometric, and described by a $T^2$ fibration. The non-classical
nature of the background arises from the patching between local
descriptions, which  entails non-trivial elements of the
T-duality group acting on the $T^2$.
Such fibrations will in general have  defects where a local description
in terms of the heterotic string on a smooth background is no longer possible. 

For concreteness, we consider the compactification of the heterotic string to six dimensions so that we have 
locally a $T^2$ fibration over a complex one-dimensional base. At certain points of the base there are
defects and our goal is to describe the low-energy dynamics living on the defects themselves.
This is achieved by dualizing the configuration to F-theory, where the
dynamics on the defects can be characterized by purely geometric means. 
Generically, the F-theory background dual to a given defect on the heterotic side are highly
singular. In some cases we are able to resolve the singularity crepantly by
performing a finite number of blow-ups in the base of the
fibration. For all the cases where this resolution is possible we
construct the resulting smooth geometry. The blow-ups correspond to
giving vevs to tensor multiplets of the 6d (1,0) theory on the defect,
such that it flows to a Lagrangian description in the IR. 
For the cases that can be resolved
the emerging theories can be related to 6d SCFTs, such as the long known theories
of small instantons on an ADE singularity \cite{Aspinwall:1997ye, Intriligator:1997dh}, or
6d SCFTs that have been recently classified \cite{Heckman:2013pva,Heckman:2015bfa,DelZotto:2014hpa}.
Actually, the resulting theories fall into configurations whose UV completions are conjectured
to be 6d little string theories (LSTs) \cite{Bhardwaj:2015xxa, Bhardwaj:2015oru, Font:2017odl}
since they have distinctive properties of LSTs \cite{Seiberg:1997zk} such as
a mass scale and T-duality upon circle compactification. 
Moreover, there is a prescription to extract 6d SCFTs embedded in the LSTs \cite{Font:2017odl}.

The paper is organized as follows:
In section \ref{sec:basics}, after recalling the basics of heterotic compactifications on $T^2$,
we review the formulation of heterotic/F-theory duality in terms of a map
between genus-two curves and K3 surfaces. Moreover, we discuss how it can be
used to study non-geometric heterotic backgrounds in terms of K3 fibered Calabi-Yau three-folds. 
In section \ref{sec:resolution}, we first explain the procedure to resolve singularities
and then apply the formalism to local heterotic degenerations which
admit a geometric description in some duality frame. We also discuss truly non-geometric 
models and describe a kind of duality between different non-geometric and geometric defects.
In section \ref{sec:catalog} we summarize the classification of all possible local heterotic models, both
geometric and non-geometric, admitting F-theory duals that can be
resolved crepantly into smooth Calabi-Yau three-folds. We end with 
some final comments.


\section{Non-geometric heterotic vacua and F-theory}\label{sec:basics}
In this section we describe what we actually mean by non-geometric heterotic string vacua. We construct them in two steps: first we compactify the ten-dimensional string theory on a two-torus; then we use the duality group of the moduli to get a non-trivial, i.e.\ non-geometric, identification of these fields when going along a loop of non-trivial homotopy.  We will also  review the duality between the heterotic string 
and F-theory \cite{Morrison:1996na,Malmendier:2014uka,Gu:2014ova}, in preparation for section~\ref{sec:resolution} where we use the F-theory representation to get a better handle 
on the low-energy degrees of freedom of the non-geometric heterotic vacua.

\subsection{Heterotic string on $T^2$}
From the compactification of the heterotic string on a torus $T^2$, we obtain the following moduli fields in eight dimensions:
\begin{itemize}
 \item A complexified K\"ahler modulus $\rho=\int_{T^2} B+\omega\wedge\bar\omega$ with $B$ the Kalb-Ramond two-form and $\omega$ the holomorphic one-form of the torus which can be obtained from the metric on $T^2$.
 \item The complex structure modulus $\tau=\int_b \omega/\int_a \omega$, where $a$ and $b$ denote the two generators of the non-trivial one-cycles of the torus.
 \item Furthermore, there are 16 complex Wilson line moduli from the Cartan generators of the non-abelian gauge group of the heterotic string, i.e.\ $\beta^i=\int_a A^i +i\int_b A^i$.
\end{itemize}
As it is well known there are dualities among torus compactifications. Therefore, the local moduli space $O(2;\mathbb R)\times O(2+n_{WL};\mathbb R)\backslash O(2,2+n_{WL};\mathbb R)$ of the heterotic $T^2$ compactification becomes the Narain space \cite{Narain:1985jj}
\begin{equation}
 O(2;\mathbb R)\times O(2+n_{WL};\mathbb R)\backslash O(2,2+n_{WL};\mathbb R)/O(2,2+n_{WL};\mathbb Z)\,,
\end{equation}
with $n_{WL}$ the number of Wilson line moduli switched on. The cases of interest to us are those with none or one non-vanishing Wilson line modulus. In these situations the 
heterotic/F-theory duality map is known explicitly and can be used to analyse the heterotic vacua.

\subsection{Vacua with varying moduli fields}

To set the ground for the second step of our compactification we rewrite the above moduli space. For $n_{WL}=1$ we can map\footnote{We should note that this map is a priori only well-defined from $\mathbb H_2/Sp(4,\mathbb Z)$ to the Narain moduli space. 
Only on $O(2;\mathbb R)\times O(3;\mathbb R)\backslash O(2,3;\mathbb R)/SO^+(2,3;\mathbb Z)$, where $SO^+(2,3;\mathbb Z)$ is an order four subgroup of $O(2,3;\mathbb Z)$,
it becomes a well-defined bijective map.} the Narain moduli space to the Siegel upper half plane of genus-two curves
\begin{equation}\label{eq:Siegel-upper-half-plane}
 \mathbb H_2=\left\{\Omega=\left(\begin{array}{cc} \tau &\beta \\\beta & \rho\end{array}\right)\Big| \Im(\det(\Omega))>0 \wedge \Im(\rho)>0\right\}
\end{equation}
quotiented by an $Sp(4,\mathbb Z)$-action
\begin{equation}\label{eq:Sp4Z-trafo}
\Omega\rightarrow (A \Omega +B)(C\Omega +D)^{-1} \qquad \textmd{with}\qquad
 \left(\begin{array}{cc} A & B \\ C & D\end{array}\right)\in Sp(4,\mathbb Z)\,. 
\end{equation}
The advantage of this rewriting is firstly that the above moduli $\rho$, $\tau$, and $\beta$ are just the entries of $\Omega$ as denoted in \eqref{eq:Siegel-upper-half-plane}. Secondly, in this representation it is natural to assign to every moduli space point $p$ a genus-two curve $\mathcal C_p$ with complex structure
\begin{equation}
 \Omega_{ij}=\int_{b_i}\omega_j\,.
\end{equation}
Here $a_i$, $b_i$, and $\omega_j$ are respectively the non-trivial one-cycles and holomorphic one-forms of $\mathcal C_p$, with 
normalization $\int_{a_i}\omega_j=\delta_{ij}$.
In this way we obtain a geometrification of our moduli.

In the vein of F-theory, we use this geometrification to construct six-dimensional heterotic string vacua with varying moduli fields. Therefore, we let the heterotic torus fibration vary adiabatically along two real dimensions or one complex dimension which we parametrize by $t$. For the moduli to fulfill the  (BPS) equations of motion they must vary holomorphically in $t$. To obtain the wanted \emph{(globally) non-geometric} configurations, we puncture the $t$-plane and allow for `$Sp(4,\mathbb Z)$-patchings' of the moduli fields when encircling the punctures, i.e.\ we identify dual theories when going along a non-contractible loop.\footnote{Non-geometric configurations usually refer to situations where the metric is identified with its inverse along a non-trivial path. However, in our case generically we have mixings of all three moduli of which $\rho\rightarrow 1/\rho$ is just a subgroup.} Since every $Sp(4,\mathbb Z)$-orbit in $\mathbb H_2$ is identified with exactly one genus-two curve, holomorphic genus-two fibrations are the natural candidates to encode such vacua. If such a fibration is non-trivial it will degenerate in complex codimension one. These degeneration points are the punctures of the $t$-plane and the kind of singularity is in one-to-one relation with a certain $Sp(4,\mathbb Z)$-duality transformation on the moduli.

Now, it is a happy coincidence that all the degenerations of genus-two curves were classified 
by Ogg-Namikawa-Ueno \cite{ogg66,Namikawa:1973yq}. 
This mathematical result gives us a huge list of non-geometric heterotic string vacua. However, to understand them we have to make sense of the singularity loci of the fibration. In string theory we are used to the appearance of new light degrees of freedom which cure the theory when we run towards a seeming singularity. Since, we do not know how to do such an analysis for these specific compactifications directly on the heterotic side, we use the duality with F-theory to    
study them. The localised physical objects, which lie at the center of the genus two-degeneration, are called \emph{$T$-fects}, which is short for $T$-duality defect.

\subsection{Mapping the setting to F-theory}

Since the invention of F-theory \cite{Vafa:1996xn} it is known that the heterotic string on $T^2$ is dual to F-theory compactified on K3 \cite{Morrison:1996na}. This duality is best understood in the large volume/stable degeneration limit \cite{Berglund:1998ej,Morrison:1996na}. This special point in moduli spaces means that on the heterotic side we have $\rho\rightarrow i \,\infty$ and on the F-theory side the K3 degenerates into two `del Pezzo nine' surfaces which intersect each other along a $T^2$. The identification of the moduli data is now as follows: $\tau$ is the complex structure of the F-theory $T^2$ at the intersection, and the Wilson lines are encoded in the intersection points (spectral cover data \cite{Friedman:1997yq}) of the respective nine exceptional curves of the two dP$_9$'s with the $T^2$. Unfortunately such a detailed identification with all Wilson line moduli non-vanishing exists only for this special point in moduli space. However, for the case which we consider in this article, i.e.\ only one Wilson line non-vanishing, we can even do better. In this case the map between the moduli fields on both sides is known along the whole moduli space \cite{Clingher:3503c,Malmendier:2014uka}.

For $n_{WL}=1$, in the $E_8\times E_8$ heterotic string, the hypersurface describing the elliptically fibered F-theory K3 takes the following form:
\begin{equation}\label{eq:K3-E7xE8}
 y^2=x^3 + (a\, u^4 v^4 + c\,u^3 v^5)\,x\,z^4 + (b\,u^6 v^6 + d\,u^5 v^7 + u^7 v^5)\,z^6
\end{equation}
where $x$, $y$, $z$ and $u$, $v$ are the homogeneous coordinates of fiber ambient variety $\mathbb P_{2,3,1}$ and the base $\mathbb P^1$, respectively. This K3 has a $II^*$ singularity at $v=0$ and a $III^*$ at $u=0$ which correspond to an $E_8$ and an $E_7$ gauge group, respectively---matching the remaining unbroken heterotic gauge group after switching on one Wilson line modulus. Furthermore, the Picard number of this manifold is 17. Therefore, its complex structure moduli space \cite{Aspinwall1996c} exactly agrees with the heterotic moduli space.\footnote{Note that this is obviously also true for the cases with $n_{WL}>1$ and one of the reasons why these two theories are dual to each other.} As mentioned already, the map between the complex structures of the 
F-theory K3 \eqref{eq:K3-E7xE8} and the heterotic moduli \eqref{eq:Siegel-upper-half-plane} is explicitly known and given by \cite{Clingher:3503c,Malmendier:2014uka}:
\begin{equation}
a=-\frac1{48}\psi_4(\Omega)\,,\quad b=-\frac1{864}\psi_6(\Omega)\,,\quad c=-4\chi_{10}(\Omega)\,, \quad d=\chi_{12}(\Omega)\,.
\end{equation}
with $\psi_4$, $\psi_6$, $\chi_{10}$, and $\chi_{12}$ the genus-two Siegel modular forms \cite{Bruinier_2008} of weight four, six, ten, and twelve, respectively. The modularity of the forms is meant with respect to the $Sp(4,\mathbb Z)$ transformation \eqref{eq:Sp4Z-trafo}.

As we pointed out already, we are interested in an understanding of the physics of the vacua given by the list of genus-two degenerations. In their classification Namikawa and Ueno \cite{Namikawa:1973yq} give the genus-two singularities explicitly in terms of fibrations of hyperelliptic curves, i.e.\ sextic equations of the form
\begin{equation}\label{eq:sextic}
 y^2=c_6(t)\,x^6+c_5(t)\,x^5+\ldots +c_1(t)\,x + c_0
\end{equation}
with the $c_i(t)$'s functions (or sections) of $t$. Furthermore, all the hyperelliptic curve fibrations are in a canonical form, in the sense that the singularity lies at $t=0$. Having the genus-two fibrations in the form of \eqref{eq:sextic} is very convenient for us because we can use the relations between the genus-two Siegel modular forms and the Igusa-Clebsch 
invariants\footnote{See appendix C of \cite{Font:2016odl} for the explicit form of the Igusa-Clebsch invariants in terms of the coefficients of the sextic.} $I_2$, $I_4$, $I_6$, $I_{10}$ of the sextic \cite{Igusa:1962},
\begin{equation}
\label{SiegelICmap}
\begin{aligned}
&I_2(c_i) = \frac{\chi_{12}(\Omega)}{\chi_{10}(\Omega)} \,,
&I_4(c_i) = 2^{-4}\cdot 3^{-2} \psi_4(\Omega) \,,\\
&I_6(c_i) = 2^{-6}\cdot 3^{-4} \psi_6(\Omega) + 2^{-4}\cdot 3^{-3}\frac{\psi_4(\Omega) \chi_{12}(\Omega)}{\chi_{10}(\Omega)} \,,
&I_{10}(c_i) = 2^{-1}\cdot 3^{-5} \chi_{10}(\Omega) \,,
\end{aligned}
\end{equation}
to write down the K3 coefficients $a$, $b$, $c$, $d$ as functions of the sextic coefficients $c_i$. In the end, we obtain for every genus-two singularity a K3 fibration over the $t$-plane with the K3 fibre degenerating at $t=0$. In the next section we will look at these F-theory singularities and try to resolve them if possible. In this way we get some insight about the objects which live at 
these six-dimensional loci.

The F-theory K3 dual to the $\mathrm{Spin}(32)/\ZZ_2$ heterotic string compactified on $T^2$ with one Wilson line is also 
known \cite{McOrist:2010jw, Malmendier:2014uka}. It is described by:
\begin{equation}
y^2=x^3+  v( u^3 + a\, u v^2 +b \, v^3) \,x^2 z^2+ v^7(c\,u+d\,v)\, x\, z^4 \,.
\label{K328a}
\end{equation}
Putting the equation into Weierstra\ss\ form and computing the discriminant shows that this K3
has singularities of type $\mathrm{I}^*_{10}$
($SO(28)$) at $v=0$, and of type $\mathrm{I}_2$ ($SU(2)$) at
$cu+dv=0$, for generic coefficients. Hence, the gauge group is
$\mathrm{Spin}(28)\times SU(2)/\ZZ_2$. When $c\equiv 0$ the group enhances to $\mathrm{Spin}(32)/\ZZ_2$.

Before we finish this section and go on to the resolutions, we should note  that the map from the F-theory side to the heterotic side is much more involved, see for instance \cite{Garcia-Etxebarria:2016ibz} for a first step into this direction.

\section{Resolution of singularities: procedure and examples}\label{sec:resolution}
Having established the duality map between the heterotic vacua and the F-theory vacua, we look now at the resolution of  the singularities. Put differently in terms of F-theory language, we move onto the tensor branch of the theory to get an insight into the degrees of freedom which lie at the heart of the genus-two degenerations.

\subsection{General strategy}
We will always work with a Weierstra\ss{} model in the following, i.e.\ the elliptic fibration is always represented in terms of a hypersurface equation of the form:
\begin{equation}\label{eq:Weierstrass-eqn}
 y^2=x^3 + f(\xi_i)\,x\,z^4 + g(\xi_i)\,z^6
\end{equation}
where $x$, $y$, $z$ are again the homogeneous coordinates of $\PP_{2,3,1}$ and $f$ and $g$ are sections of some line bundles over the base $B\ni \xi_i$. For the elliptic fibration to be Calabi-Yau the line bundles of $f$ and $g$ have to be $K^{-4}_B$ and $K^{-6}_B$, respectively, with $K_B$ the canonical bundle of the base.

Throughout this article, we will call a singularity resolved if the elliptic curve has only minimal singularities \cite{kodaira123} (or Kodaira type singularities) along the base, i.e.\ there are no points along the discriminant locus of \eqref{eq:Weierstrass-eqn} where $f$ vanishes to order four or higher and simultaneously $g$ to order six or higher.  Furthermore, in our examples the base on the F-theory side is given by a (trivial) $\PP^1$ fibration over the $t$-plane.
Since  the coefficient in front of the $u^7\,v^5$ term in \eqref{eq:K3-E7xE8} is constant, there is no such non-minimal singularity along $v=0$. Therefore, we only have to look at the $u$-$t$-patch for such points and, as it turns out, in the beginning there is just one non-minimal singularity namely at $u=t=0$. To get rid of this non-minimal point we follow \cite{Aspinwall:1997ye} and blow-up the base at this point. However, we do this in a rather toric manner by introducing the maximal amount of crepant\footnote{Crepant in the sense that the proper transform of  the hypersurface equation \eqref{eq:K3-E7xE8} after the base blow-up is still Calabi-Yau. We do not claim that the canonical class of the base does not change which is obviously wrong.} blow-ups at once at the non-minimal point and not in a blow-up after blow-up procedure. 
Afterwards we search for non-minimal points along the newly introduced exceptional curves and, if necessary, apply our `toric blow-up procedure' again.
The analysis can also be applied
to the dual F-theory K3 of the $\mathrm{Spin}(32)/\ZZ_2$ heterotic string, described by the equation \eqref{K328a}.

\subsubsection*{Toric blow-up procedure}

As a first step, we choose local affine coordinates $\xi_i$ on $B$ such that the  non-minimal singularity lies at $\xi_1=\xi_2=0$. We expand the sections $f$ and $g$ in these coordinates,
\begin{equation}\label{eq:f_and_g-expansion}
f=\sum_i f_i \,\xi_1^{m^1_i}\xi_2^{m^2_i}\,,\qquad g=\sum_i g_i \,\xi_1^{l^1_i}\xi_2^{l^2_i}\,,
\end{equation}
and collect the minimal exponents $\mathbf m_i$ and $\mathbf l_i$. 
Next we look for all `toric' blow-up \cite{fulton1993introduction} directions $\mathbf n_j$ which are crepant.  
For the elliptic fibration to remain Calabi-Yau, the blow-up $\mathbf n$ must involve the fibre coordinates $x$ and $y$ too:
\begin{equation}\label{eq:blow-up}
\xi_1,\, \xi_2,\,x,\,y\quad \mapsto \quad e^{n_1}\,\tilde\xi_1,\,e^{n_2}\,\tilde \xi_2,\,e^{2(n_1+n_2-1)}\,\tilde x,\,e^{3(n_1+n_2-1)}\,\tilde y\,.
\end{equation}
Hence, the canonical class of the ambient variety after the blow-up is given by $E$ times the last column in the following table:
\begin{equation}
\begin{array}{|c|c|c|c|c|c|c|}
\hline
     & \xi_1 & \xi_2 & x & y & e & \sum\\\hline
 E & n_1 & n_2 & 2(n_1+n_2-1) & 3(n_1+n_2-1) & -1 &  6(n_1+n_2-1)\\ \hline
\end{array}\,,
\end{equation}
where $-E$ is the divisor class of the exceptional divisor $e=0$. Since we demand that our resolution of the Weierstra\ss{} equation is crepant, $e^{6(n_1+n_2-1)}$ must factor off the hypersurface equation \eqref{eq:Weierstrass-eqn} when we take its proper transform after applying \eqref{eq:blow-up}. This amounts then to the constrains
\begin{equation}\label{eq:allowed-blow-up-directions}
(m^1_i-4)n_1+(m^2_i-4)n_2=:\tilde{\mathbf m}_i\cdot \mathbf n\ge -4 \quad \textmd{and}\quad (l^1_i-6)n_1+(l^2_i-6)n_2=:\tilde{\mathbf l}_i\cdot \mathbf n\ge -6 
\end{equation}
which must be fulfilled for all $\tilde{\mathbf m}_i$ and $\tilde{\mathbf l}_i$. The solutions to these inequalities is the set of toric blow-ups that we introduce.

After this resolution step, we have to check whether there are no non-minimal points along the just introduced exceptional curves. If there are any of them, we have to repeat the procedure at these points. We are done if we got rid of all  the non-Kodaira type singularities.

\subsection{Geometric models: small instantons on ADE singularities}

Throughout this article, we study configurations which already cancel their NS5-charge locally, i.e.\ $d\, H_3 \equiv 0$. Hence, 
\begin{equation}
\label{local_cancellation}
 \int_{B_4} d\, H_3 = \int_{\partial B_4} H_3 = 0 \, ,
\end{equation}
where $B_4$ denotes the $T^2$-fibration over the disc $D_2$ with the singularity of the fibration at its center. 
Now, the modified Bianchi identity for $H_3$ reads
\begin{equation}\label{modified_bianchi_id}
d\, H_3 = \tfrac{\alpha'}4 \left({\rm tr} F^e\wedge F^e - {\rm tr} F^A\wedge F^A  \right) \, ,
\end{equation}
with $F^e$ and $F^A$ the curvature of the spin bundle and gauge bundle, respectively. Therefore, the three-form flux can be written as  
\mbox{$H_3 =  d\, B_2 - \tfrac{\alpha'}4 (\omega_3^A - \omega_3^e)$}, where $\omega_3^A$ and $\omega_3^e$ are the Yang-Mills and Lorentz Chern-Simons forms.  
Since the boundary of $B_4$ is a $T^2$-fibration over an $S^1$ encircling the singularity, and due to \eqref{modified_bianchi_id}, the 
equation \eqref{local_cancellation} may be rewritten as\footnote{Let us note here that $B_2$ is not an ordinary two-form but rather a gauge field otherwise the first term on the right-hand side of \eqref{bianchi_id_integrated} would vanish trivially.}
\begin{equation}\label{bianchi_id_integrated}
 \int_{\partial B_4} d\, B_2 - \tfrac{\alpha'}4 (\omega_3^A - \omega_3^e) =  
 \int_{T^2} B_2 \Bigg|_0^{2\pi} + \tfrac{\alpha'}4 \int_{\partial B_4} \omega_3^e - \tfrac{\alpha'}4\int_{B_4} {\rm tr} F^A\wedge F^A \,.
\end{equation}
The first and second term on the right-hand side of \eqref{bianchi_id_integrated} encode the shift in the real part of $\rho$ and $\tau$, respectively, and the last term is just the instanton number. Hence, a monodromy in $\rho$ and $\tau$ ---if not cancelled between them \cite{Garcia-Etxebarria:2016ibz}---has to be compensated by small instantons localized at the singularity. 

Now we want to consider, in this section, resolutions of heterotic models 
which on the genus-two side have a Namikawa-Ueno (NU) degeneration $[\mathrm{I}_{n-p-0}]$,
$[\mathrm{I}_n-\mathrm{I}^*_p]$ and $[\mathrm{K}-\mathrm{I}_n]$, with $\mathrm{K}=\mathrm{II}^*, \mathrm{III}^*, \mathrm{IV}^*$  \cite{Namikawa:1973yq}.
Here we use the notation $\kod{[K_1-K_2-0]}\equiv \kod{[K_1-K_2]}$ for the Namikawa-Ueno degenerations.
Based on the monodromy action on the moduli and the modified Bianchi identity \eqref{modified_bianchi_id}, these models are expected to describe heterotic compactifications with small instantons sitting at
ADE singularities. For example, in the $[\mathrm{II}^{\ast}-\mathrm{I}_n]$ model the monodromy is
\begin{equation}
\label{monoe8}
\tau \rightarrow -\frac{1}{1+\tau} \, ,\quad \rho \rightarrow \rho + n - \frac{\beta^2}{1+\tau}
\, ,\quad \beta \rightarrow
\frac{\beta}{1+\tau}\, .
\end{equation}
When the Wilson line value $\beta$ is turned off, this is
precisely the monodromy of a $\kod{II^{\ast}}$ type fiber of the $\tau$ fibration. 
In general, in $[\mathrm{K}-\mathrm{I}_n]$ models
it follows that there is a number $k=\mu(c)$ of small instantons on top of the $\mathrm{K}$-type singularity.

The starting point is the genus-two model given in the NU classification. The next step is to compute the Igusa-Clebsch 
invariants that determine the $a$, $b$, $c$, and $d$ coefficients entering in the dual K3 on the F-theory side. 
In table \ref{tab:adetype} we collect the defining equations of the ADE NU models together with the vanishing degrees of the coefficients $a$, $b$, $c$, and $d$. On the F-theory side there are points where the vanishing orders of $f$, $g$, and $\Delta$ in the Weierstra\ss \ model
are non-minimal, so we proceed to resolve as explained in the preceding section.

\begin{table}[h!]
{\small
\begin{center}
\renewcommand{\arraystretch}{1.5}
\begin{tabular}{|M{1cm}|M{1.5cm}|M{5.5cm}|c|c|c|c|}
\hline
sing. & NU   type       &  local model &  $\mu(a)$ & $\mu(b)$ &$\mu(c)$& $\mu(d)$        \\[3pt]
\hline
$\mathrm{A}_{p-1}$ &  $[\mathrm{I}_{n-p-0}]$  &$ \left(t^n+x^2\right) \left(t^p+(x-\alpha)^2\right)\left(x-1\right)  $& $0$&  $0$&$n+p$ & $n+p$  
  \\[3pt]\hline
$\mathrm{D}_{p+4}$ & $[\mathrm{I}_n-\mathrm{I}^*_p]$  &$\left(t^n+(x-1)^2\right) \left(t^{p+2}+x^2\right)\left(x+t\right)  $& $2$&  $3$&$6+n+p$ & $6+n+p$  
  \\[3pt]\hline
$\mathrm{E}_6$ & $[\mathrm{IV}^{\ast}-\mathrm{I}_n]$  &$\left(t^{4}+x^3\right) \left(t^n+(x-1)^2\right) $& $4+n$&  $4$&$8+n$ & $8+n$  
  \\[3pt]\hline
$\mathrm{E}_7$ & $[\mathrm{III}^{\ast}-\mathrm{I}_n]$  &$x\left(t^{3}+x^2\right) \left(t^n+(x-1)^2\right) $& $3$&  $6+n$&$9+n$ & $9+n$  
  \\[3pt]\hline
$\mathrm{E}_8$ &   $[\mathrm{II}^{\ast}-\mathrm{I}_n]$  &$\left(t^{5}+x^3\right) \left(t^n+(x-1)^2\right) $& $5+n$&  $5$&$10+n$ & $10+n$  
  \\[3pt]\hline
\end{tabular}
\caption{Genus-two models for ADE singularities.}
  \label{tab:adetype}\end{center}
  }\end{table}

As mentioned before, in general the resolution consists of a series of base blow-ups. Each curve
is characterized by an integer equal to minus its self-intersection number,\footnote{
After the resolutions, the curve $t=0$ has self-intersection $-1$. 
We label this curve by $1^*$, instead of $1$, to indicate that it existed already before the base blow-ups.
}
and by the gauge algebra factor it supports.
This algebra is identified after checking for the presence of monodromies
following the formalism of \cite{Grassi:2011hq}. In order to determine the matter content it is also important
to give the intersection pattern of the blow-ups. Applying the resolution procedure, we obtain all these data.
The results match those obtained in  \cite{Aspinwall:1997ye}. Below we present two typical examples where we consider both
the $E_8 \times E_8$ and the $\mathrm{Spin}(32)/\ZZ_2$ heterotic string.

\subsection*{$\kod{[II^{\ast}-I_n]}$ model and $\mathrm{E}_8$ singularity}\label{ss:InE8}

The pattern of curves and self-intersection numbers is efficiently determined using the
toric geometry techniques reviewed in the preceding section. For the $E_8 \times E_8$ heterotic we find:
\begin{align}
\label{resoInE8}
&
\footnotesize{
\begin{tabular}{|cccccccccc|}
\hline
& & $\mathfrak{sp}(1)$ &$\mathfrak{g}_2$& & $\mathfrak{f}_4$& &  $\mathfrak{g}_2$& $\mathfrak{sp}(1)$ &  \\
1 & 2& 2& 3& 1& 5 & 1 & 3 & 2 & 2 \\
\hline
\end{tabular}\, 
}
\footnotesize{
\begin{tabular}{|cccccccccccc|}
\hline
 & 1 &  & & & & & &  &  &  & \\
 & $\mid$ &  & & & & & &  &  &  &  \\
 & $\mathfrak{e}_8$& & & $\mathfrak{sp}(1)$ &$\mathfrak{g}_2$& & $\mathfrak{f}_4$& &  $\mathfrak{g}_2$& $\mathfrak{sp}(1)$ &  \\
1 & 12 & 1 & 2& 2& 3& 1& 5 & 1 & 3 & 2 & 2 \\
\hline
\end{tabular}\, 
\times  }   \nonumber  \\ 
& 
\footnotesize{
\begin{tabular}{|cccccccccccc|}
\hline
 & $\mathfrak{e}_8$& & & $\mathfrak{sp}(1)$ &$\mathfrak{g}_2$& & $\mathfrak{f}_4$& &  $\mathfrak{g}_2$& $\mathfrak{sp}(1)$ &  \\
1 & 12 & 1 & 2& 2& 3& 1& 5 & 1 & 3 & 2 & 2 \\
\hline
\end{tabular}^{\, \oplus (n-1)}
\hspace*{-5mm}  \times} \\ \nonumber
& 
\footnotesize{ \times
\begin{tabular}{|ccccccccccccc|}
\hline
 & $\mathfrak{e}_8$& & & $\mathfrak{sp}(1)$ &$\mathfrak{g}_2$& & $\mathfrak{f}_4$& &  $\mathfrak{g}_2$& $\mathfrak{sp}(1)$ &  &\\
1 & 12 & 1 & 2& 2& 3& 1& 5 & 1 & 3 & 2 & 2 & $1^*$\\
 & $\mid$ &  &  &  &  &  &   &   &   &   &   &\\
  & 1         &  &  &  &  &  &   &   &   &   &   &\\
\hline
\end{tabular}\, 
\, . }
\end{align}
This result agrees with the theory of $k=10+n$, with $n\ge 1$, pointlike
instantons on the $\mathrm{E}_8$ singularity as given in \cite{Aspinwall:1997ye}.
Deleting the node associated to $t=0$, gives the tensor branch description of a 6d SCFT embedded in the LST. 
This is the situation which was implicitly assumed in \cite{Font:2016odl}.

Starting from the $\mathrm{Spin}(32)/\ZZ_2$ heterotic string the pattern of gauge factors and self-intersection numbers turns out to be: 
\begin{equation}
\footnotesize{
\label{newE8}
\begin{tabular}{|cccccccc|}
\hline
 & & & & & $\mathfrak{sp}(3k\text{-}32)$ & & \\
 & & & & & 1 & & \\
 & & & & & $\mid$ & & \\
$\mathfrak{sp}(k)$ &  $\mathfrak{so}(4k\text{-}16)$ &$\mathfrak{sp}(3k\text{-}24)$& $\mathfrak{so}(8k\text{-}64)$&
  $\mathfrak{sp}(5k\text{-}48)$&$\mathfrak{so}(12k\text{-}112)$&$\mathfrak{sp}(4k\text{-}40)$&
 $\mathfrak{so}(4k\text{-}32) $ \\ 
1*&4&1&4&1&4&1&4 \\\hline
\end{tabular} \, .
}
\end{equation}
The first factor $\small{\mathfrak{sp}(k)}$ arises from the singularity at $t=0$ which, before the base blow-ups, is non-minimal only at $u=t=0$. 
The total number of base blow-ups is eight.  
Notice that the structure of the intersections 
conforms to the extended Dynkin diagram of $E_8$, in agreement with the analysis in \cite{Blum:1997mm}. 
Dropping the node corresponding to $t=0$, gives the tensor branch of a 6d SCFT embedded in the LST, with the $\mathfrak{sp}(k)$
remaining as a flavor symmetry. 

\subsection*{$[\kod I_0^{\ast}-\kod I_n]$ model and $\mathrm{D}_4$ singularity}

In the $E_8 \times E_8$ case the resolution gives
\begin{equation}
\label{singD4}
\begin{tabular}{|ccccc|}
\hline
 &&   $\mathfrak{sp}(1)$ & $\mathfrak{g}_2$& \\
1 &2&   2 &3  & 1 \\\hline
\end{tabular} \, 
\begin{tabular}{|cc|}
\hline 
$\mathfrak{so}(8)$& \\
4&1 \\\hline
\end{tabular}^{\, \oplus (n-1)}\, 
\begin{tabular}{|cccc|}
\hline
 $\mathfrak{g}_2$& $\mathfrak{sp}(1)$& &\\
3& 2 &2 & $1^*$\\\hline
\end{tabular} \, .
\end{equation}
This result agrees with the theory of $k=6+n$, $n\ge 1$, pointlike
instantons on the $\mathrm{D}_4$ singularity obtained in \cite{Aspinwall:1997ye}. When $n=0$ we instead find
\begin{equation}
\label{singD40}
\begin{tabular}{|ccccccc|}
\hline
 &&  $\mathfrak{sp}(1)$ &$\mathfrak{g}_2$&$\mathfrak{sp}(1)$ & &\\
1 & 2&     2 &2&2&2 & $1^*$\\\hline
\end{tabular} \, .
\end{equation}

For the $\mathrm{Spin}(32)/\ZZ_2$ heterotic string, the resolution leads to
\begin{equation}
\label{modD4}
\begin{tabular}{|cc|}
\multicolumn{2}{c}{$n=0$}\\
\hline
$\mathfrak{sp}(6)$ &  $\mathfrak{so}(7)$\\ 
1*&1\\\hline
\end{tabular} \  ,  \hspace*{1cm} 
\begin{tabular}{|cc|}
\multicolumn{2}{c}{$n=1$}\\
\hline
$\mathfrak{sp}(7)$ &  $\mathfrak{so}(12)$\\ 
1*&1\\\hline
\end{tabular} \ , \hspace*{1cm} 
\begin{tabular}{|ccc|}
\multicolumn{3}{c}{$n\ge 2$}\\
\hline
& $\mathfrak{sp}(k\text{-}8)$ & \\
& 1 & \\
& $\mid$ & \\
$\mathfrak{sp}(k)$ & $\mathfrak{so}(4k\text{-}16)$ & $\mathfrak{sp}(k\text{-}8)$ \\ 
1*&4&1\\
& $\mid$ & \\
& 1 & \\
& $\mathfrak{sp}(k\text{-}8)$ & \\
\hline
\end{tabular}\, .
\end{equation}
The number of blow-ups is one for $n=0,1$ and four for $n\ge 2$.

\subsection{Non-geometric models and duality web}
\label{sec:dual}

As seen in the previous section, in models
corresponding to small instantons on ADE singularities,
the explicit formulation of heterotic/F-theory
duality in terms of the map between genus-two and K3 fibrations
confirms the results expected from the monodromies of the moduli fields.
We now turn to heterotic models with monodromies which
are non-geometric in all T-duality frames. This is the most interesting
situation, since a priori it is not clear if such degenerations are
allowed. 

A simple example of a non-geometric degeneration is the Namikawa-Ueno
$\kod{[III-III]}$ singularity which has monodromy:
\begin{equation}
\label{ex33}
\tau \rightarrow \frac{\rho}{\beta^2-\rho\tau} \, ,\quad \rho
\rightarrow \frac{\tau}{\beta^2-\rho\tau} \, ,\quad \beta
\rightarrow -\frac{\beta}{\beta^2-\rho \tau} \, .
\end{equation} 
When $\beta = 0$ we obtain a ``double elliptic''
fibration which encircling the heterotic
degeneration produces the monodromy $\tau\to-1/\tau \, , \rho
\to- 1/\rho $. 
The equation for the hyperelliptic curve for the $\kod{[III-III]}$  singularity is:
\begin{equation}\label{eq:hyperell_III-III}
y^2 = x  (x - 1)  (x^{2} + t)  \left[(x-1)^2 + t\right]\, .
\end{equation}
Applying the resolution procedure gives the same six-dimensional theory as $\kod{[I_0-I_0^{\ast}]}$, 
c.f. \eqref{singD40} and \eqref{modD4}, namely the theory of six small instantons on a $\mathrm{D}_4$ singularity.
We have found that, in several non-geometric models of type 2 in the NU
list, the dual CY admits a smooth resolution and, moreover, the resulting low
energy physics is  described by the theory of small instantons on ADE singularities.

\begin{table}[t]\begin{center}
\renewcommand{\arraystretch}{1.5}
\begin{tabular}{|c|c|}
\hline
$\mu(c)$ & dual models   \\
\hline \hline
4& $\kod{[I_0-IV]}
  \, ,  \,\kod{[II-II]}$  \\ \hline 
5&$[\kod{IV}-\kod{I}_1] \, , \,\kod{[II-III]}$ \\ \hline 
6& $\kod{[I_0-I_0^{\ast}]}\, , \,
  \kod{[III-III]}\, , \,\kod{[IV-II]}$ \\ \hline 
 7& $[\kod{I}_0^{\ast}-\kod{I}_1]\, , \,\kod{[IV-III]}$ \\ \hline 
8& $\kod{[I_0-IV^{\ast}]}\, ,
  \,\kod{[I_0^{\ast}-II]}$ \\ \hline 
9& $\kod{[I_0-III^{\ast}]} \,
  ,  \,\kod{[I_0^{\ast}-III]}$ \\ \hline 
10& $\kod{[I_0-II^{\ast}]} \, ,\,
  \,\kod{[I_0^{\ast}-IV]}$ \\ \hline 
11& $\kod{[II-III^{\ast}]} \, ,\,
  \kod{[IV^{\ast}-III]}$ 
\\ 
\hline 
\end{tabular}
\caption{Dual models: the NU degenerations in the same row give rise
  to the same theories after resolution of the dual F-theory model.}
  \label{tab:dual}\end{center}\end{table}

As explained in \cite{Font:2016odl}, models with the same resolution such as
$\kod{[III-III]}$ and $\kod{[I_0-I_0^{\ast}]}$, can be related by certain duality moves.
As a rule, such dual models appear when the sum of the vanishing orders of the
discriminant for their two Kodaira components, or equivalently the
vanishing order  $\mu(c)$, is the same. In table \ref{tab:dual} we display
all the models satisfying this condition and admitting dual
smooth Calabi-Yau resolutions. For all the models in table \ref{tab:dual} we
explicitly performed the F-theory resolution. 
For both heterotic strings, we verified that for all the degenerations in a row the same theory arises.
The $\kod{[IV^{\ast}-II]}$ model was originally included among the duals at $\mu(c)=10$. However, in the 
$\mathrm{Spin}(32)/\ZZ_2$ heterotic string its resolution differs from that of $\kod{[I_0-II^{\ast}]}$ and closer inspection
shows that this is also the case in the $E_8\times E_8$ heterotic string. Nonetheless, the theories could be connected 
by RG flow \cite{Heckman:2013pva,Heckman:2015ola}. A similar situation arises for the $ \kod{[IV-IV]}$ 
model at $\mu(c)=8$  \cite{Font:2017odl}. 

\section{A catalog of T-fects}\label{sec:catalog}

In this section, we summarize our findings for the Namikawa-Ueno models for which 
we could construct the dual CY resolution.  Altogether there is a total of 49 
models out of the 120 entries in the NU classification.
The resulting patterns in the  $E_8 \times E_8$ case were thoroughly reported in \cite{Font:2016odl} and for 
the $\mathrm{Spin}(32)/\ZZ_2$ heterotic string they appeared in \cite{Font:2017odl}. 
Here we only present a few examples in both heterotic string theories.

\subsection{Elliptic type 1}

The elliptic type 1 NU degenerations are characterized by a monodromy action
that mixes the three moduli. Even though the corresponding
heterotic models lack a geometric interpretation, the dual F-theory
resolutions are similar to those discussed in the previous section.
In table \ref{tab:elliptic1} we gather the models whose F-theory duals admit 
a smooth CY resolution. For example, the resolution
for the $\kod{[IX-1]}$ degeneration in the $E_8 \times E_8$ heterotic is given by
\begin{equation}
\begin{tabular}{|ccccccccccccccccc|}
\hline
 &  $\mathfrak{su}(2)$ &$\mathfrak{so}(7)$& $\mathfrak{su}(2)$&
  &$\mathfrak{e}_7$&&&$\mathfrak{sp}(1)$& $\mathfrak{g}_2$& &$\mathfrak{f}_4$ && $\mathfrak{g}_2$ & $\mathfrak{sp}(1)$&&\\
1 & 2&     3 &2&1&8&1&2&2&3&1&5&1&3&2&2&1* \\\hline
\end{tabular} \, .
\end{equation}
In the $\mathrm{Spin}(32)/\ZZ_2$ we find
\begin{equation}
\label{modIX}
\begin{tabular}{|ccccccc|}
\hline
$\mathfrak{sp}(8)$ &  $\mathfrak{so}(20)$ &$\mathfrak{sp}(4)$& $\mathfrak{so}(12)$& &
  $\mathfrak{su}(2)$&$\mathfrak{so}(7)$ \\ 
1*&4&1&4&1&2&3 \\\hline
\end{tabular} \, .
\end{equation}

\begin{table}[h!]\begin{center}
\renewcommand{\arraystretch}{1.5}
\begin{tabular}{|c|c|c|c|c|}
\hline
NU model & $\mu(a)$ & $\mu(b)$ & $\mu(c)$ & $\mu(d)$  \\
\hline \hline
$\kod{[I_{0-0-0}]}$  &0& 0&0&0 \\ \hline 
$\kod{[V]}$  &2& 3&5&6 \\ \hline 
$\kod{[VII]}$ &2& 3&5&6 \\ \hline 
 $\kod{[VIII-1]} $&$\infty$& $\infty$&4&$\infty$ \\ \hline 
$\kod{[IX-1]}$ &$\infty$& $\infty$&8&$\infty$  \\\hline 
\end{tabular}
\caption{Elliptic type 1 models.}
  \label{tab:elliptic1}\end{center}\end{table}

\subsection{Elliptic type 2}

Type 2 models in the NU list comprise degenerations of type
$[\mathrm{K}_1-\mathrm{K}_2-m]$, with $m \geq 0$, where $\mathrm {K}_1$ and $\mathrm {K}_2$  are one of the Kodaira type singularities
for the two genus-one components of the genus-two surface $\Sigma$, plus additional sporadic
models denoted as $[2\mathrm{K}-m]$ and
$[\mathrm{K}_1-\mathrm{K}_2-\alpha]$. None of the latter,
nor any of the models with $m\neq0$, lead to a dual CY admitting a  smooth crepant resolution.
We find 20 models that can be resolved. They are displayed in
table \ref{tab:elliptic2} using again the notation $[\mathrm{K}_1-\mathrm{K}_2-0]\equiv
[\mathrm{K}_1-\mathrm{K}_2] $.
The models of type $\kod{[I_0-K_2]}$ correspond to a configuration of
$k=\mu(c)$ pointlike instantons on the $\kod{K_2}$ singularity. The remaining models are non-geometric since
their monodromies involve  non-trivial actions on the torus volume. 
However, as discussed in section \ref{sec:dual}, many of these
models lead to the same resolutions as the geometric ones. 

\begin{table}[t]\begin{center}
\renewcommand{\arraystretch}{1.5}
\begin{tabular}{|c|c|c|c|c||c|c|c|c|c|}
\hline
NU model & $\mu(a)$ & $\mu(b)$ & $\mu(c)$ & $\mu(d)$&NU model & $\mu(a)$ & $\mu(b)$ & $\mu(c)$ & $\mu(d)$  \\
\hline \hline
$\kod{[I_0-I_0]}$  &0&0&0&0 &$\kod{[II-IV]}$ &3& 3&6&6 \\ \hline 
  $\kod{[I_0-II]}$ &1& 1&2&2 &$\kod{[I_0^{\ast}-II]}$ &3& 4&8&8 \\ \hline 
 $\kod{[I_0-III]} $&1& 2&3&3 &$\kod{[II-IV^{\ast}]}$ &5& 5&10&10 \\ \hline 
$\kod{[I_0-IV]}$ &2& 2&4&4  &$\kod{[II-III^{\ast}]}$ &4& 7&11&11 \\\hline 
$\kod{[I_0-I_0^{\ast}]}$ &2& 3&6&6  &$\kod{[III-III]}$ &2& 4&6&6 \\\hline 
$\kod{[I_0-IV^{\ast}]}$ &4& 4&8&8  &$\kod{[IV-III]}$ &3& 4&7&7 \\\hline 
$\kod{[I_0-III^{\ast}]}$ &3& 6&9&9  &$\kod{[I_0^{\ast}-III]}$ &3& 5&9&9 \\\hline 
$\kod{[I_0-II^{\ast}]}$ &5& 5&10&10 &$\kod{[IV^{\ast}-III]}$ &5& 6&11&11  \\\hline 
$\kod{[II-II]}$ &2& 2&4&4 &$\kod{[IV-IV]}$ &4& 4&8&8  \\\hline 
$\kod{[II-III]}$ &2&3&5&5 &$\kod{[I_0^{\ast}-IV]}$ &4& 5&10&10 \\\hline 
\end{tabular}
\caption{Elliptic type 2 models.}\label{tab:vanishing-orders_ell-type-2}
  \label{tab:elliptic2}\end{center}\end{table}

\subsection{Parabolic type 3}

In this class there are additional models for which the monodromy factorizes into
the product of two monodromies---one Kodaira type for each handle of 
$\Sigma$---one of which is either $\mathrm{I}_n$ or $\mathrm{I}_n^{\ast}$ (the only
parabolic elements in the Kodaira list) and the other is of elliptic type. There are also 
models labeled $[\mathrm{K}_1-\mathrm{II}_{n}]$ or $[\mathrm{K}_1-\mathrm{II}_{n}^*]$
that mix all moduli but have a Kodaira type $\kod{K_1}$ monodromy for $\tau$. 

\begin{table}[h!]\begin{center}
\renewcommand{\arraystretch}{1.5}
\begin{tabular}{|c|c|c|c|c||c|c|c|c|c|}
\hline
NU model & $\mu(a)$ & $\mu(b)$ & $\mu(c)$ & $\mu(d)$&NU model & $\mu(a)$ & $\mu(b)$ & $\mu(c)$ & $\mu(d)$  \\
\hline \hline
$[\kod I_{n-0-0}]$  &0&0&$n$&$n$ &
 $[\kod{II}-\kod I_n] $&$1+n$& 1&$2+n$&$2+n$ \\ \hline 
$[\kod{III}-\kod I_n]$ &1& $2+n$&$3+n$&$3+n$  &
$[\kod{III}-\kod{II}_n]$ &1& $2+n$&$3+n$&$4+n$ \\ \hline 
$[\kod{IV}-\kod{I}_n]$ &$2+n$& 2&$4+n$&$4+n$  &
$[\kod{IV}-\kod{II}_n]$ &$2+n$& 2&$4+n$&$5+n$ \\ \hline 
$[\kod{II}_{n-0}]$ &2& 3&$5+n$&$6+n$ &
& & & & \\ \hline 
$[\kod I_n- \kod I_0^{\ast}]$ &2& 3&$6+n$&$6+n$  &
$[\kod{I}_0-\kod I_n^{\ast}]$ &2&3&$6+n$&$6+n$ \\ \hline
$[\kod{IV}^{\ast}-\kod I_n]$ &$4+n$& 4&$8+n$&$8+n$ &
$[\kod{II}-\kod I_n^{\ast}]$ &3&4&$8+n$&$8+n$ \\ \hline
$[\kod{III}^{\ast}-\kod I_n]$ &3& $6+n$&$9+n$&$9+n$ &
$[\kod{III}-\kod I_n^{\ast}]$ &3& 5&$9+n$&$9+n$ \\ \hline 
$[\kod{II}^{\ast}-\kod I_n]$ &$5+n$& 5&$10+n$&$10+n$ &
$[\kod{IV}-\kod I_n^{\ast}]$ &4& 5&$10+n$&$10+n$ \\ \hline 
$[\kod{IV}^{\ast}-\kod{II}_n]$ &$3+n$& 4&$7+n$&$9+n$ &
$[\kod{II}-\kod{II}_n^{\ast}]$ &$3+n$& 4&$7+n$&$9+3n$  \\ \hline 
$[\kod{III}^{\ast}-\kod{II}_n]$ &3& $5+n$&$8+n$&$11+n$  &
$[\kod{III}-\kod{II}_n^{\ast}]$ &3& $5+n$&$8+n$&$10+2n$ \\ \hline 
\end{tabular}
\caption{Parabolic type 3 models.}
  \label{tab:parabolic3}\end{center}\end{table}

The 19 models that can be resolved are listed in table \ref{tab:parabolic3}. 
They admit a resolution for all $n$. The
models of type $[\kod{I}_n-\kod K_2]$ or $[\kod{K}_1-\kod{I}_n]$ again correspond to
$k=\mu(d)$ pointlike instantons on the $\kod{K}_i$ singularity.
In this class we also discover dual models. Concretely, starting with the fifth row in table \ref{tab:parabolic3}, the
models in the same row have the same resolution. We illustrate the resolutions in this class with the
$[\kod{IV}^{\ast}-\kod{II}_n]$ model. In the $E_8\times E_8$ heterotic string we obtain
\begin{equation}
\label{resoIInIVstar}
\begin{tabular}{|ccccc|}
\hline
 & $\mathfrak{su}(2)$ &$\mathfrak{so}(7)$& $\mathfrak{su}(2)$& \\
1 & 2& 3 & 2 & 1\\
\hline
\end{tabular}\, 
\begin{tabular}{|cccc|}
\hline
  $\mathfrak{e}_6$&  &$\mathfrak{su}(3)$& \\
6&1 &3&1\\\hline
\end{tabular}^{\, \oplus n} \,
\begin{tabular}{|cccccc|}
\hline
  $\mathfrak{f}_4$&&$\mathfrak{g}_2$&$\mathfrak{sp}(1)$&&\\
5 &1&3&2&2&1*\\\hline
\end{tabular} \, .
\end{equation}
In the $\mathrm{Spin}(32)/\ZZ_2$ heterotic string we deduce
\begin{equation}
\label{modIVsIIn}
\begin{tabular}{|ccccc|}
\hline
$\mathfrak{sp}(n\text{+}7)$ &  $\mathfrak{so}(4n\text{+}16)$ &$\mathfrak{sp}(3n\text{+}1)$& 
$\mathfrak{su}(4n\text{+}2)$& $\mathfrak{su}(2n\text{+}2)$ \\ 
1*&4&1&2&2 \\\hline
\end{tabular} \, .
\end{equation}

\subsection{Parabolic type 4}

This class includes degenerations associated to parabolic Kodaira
singularities for both the genus-one components of $\Sigma$, of type
$[\kod K_1-\kod K_2-m]$ with $\kod{K_{1,2}} = \kod{I}_n,
\kod{I}_n^{\ast}$, plus additional degenerations of type
$[2\kod{K}_1-m]$, $[\kod{II}_{n-p}]$, and $[\kod{III}_n]$. Only
the three models shown in table \ref{tab:parabolic4} admit a dual smooth resolution. 
For instance, the resolution of $[\kod{II}_{n-3}]$, 
$n>3$,  singularity  for the $E_8 \times E_8$ heterotic string reads
{\small 
\begin{align}
\label{resoIInp2}
&
\begin{tabular}{|ccccccccccc|}
\hline
&  $\mathfrak{su}(2)$ &$\mathfrak{so}(7)$&& $\mathfrak{so}(9)$&$\mathfrak{sp}(1)$& $\mathfrak{so}(11)$ &$\mathfrak{sp}(2)$&$\mathfrak{so}(13)$& $\mathfrak{sp}(3)$\\
1 &      2 &3&1&4 &  1&4& 1& 4& 1 \\\hline
\end{tabular}\,
\begin{tabular}{|cc|}
\hline
$\mathfrak{so}(14)$& $\mathfrak{sp}(3)$\\
4& 1
\\\hline
\end{tabular}^{\, \oplus (n-4)} \hspace*{-5mm}\times \nonumber\\
& \times 
\begin{tabular}{|cccccccccc|}
\hline
$\mathfrak{so}(13)$&$\mathfrak{sp}(2)$ &$\mathfrak{so}(11)$  
  & $\mathfrak{sp}(1)$&  $\mathfrak{so}(9)$ &&$\mathfrak{g}_2$&$\mathfrak{sp}(1)$&&\\
4& 1& 4&  1 & 4&1&3&2&2&1*
\\\hline
\end{tabular} \, .
\end{align}
}
This result is similar to the resolution of  $k > 12$  instantons on a $D_7$ singularity \cite{Intriligator:1997dh}.
Such similarity is also observed in the $\mathrm{Spin}(32)/\ZZ_2$ heterotic string. 
For example, in the case of the same $[\kod{II}_{n-3}]$, $n>3$, degeneration we obtain
\begin{equation}
\label{modIIn3}
\begin{tabular}{|cccccc|}
\hline
 &  $\mathfrak{sp}(n\text{+}2)$ & &  & & \\
 & 1  & &  & & \\
  &  $\mid$ & &  & & \\
$\mathfrak{sp}(n\text{+}8)$ &  $\mathfrak{so}(4n\text{+}20)$ &$\mathfrak{sp}(2n\text{+}2)$& $\mathfrak{so}(4n\text{+}4)$&
  $\mathfrak{sp}(2n\text{-}6)$&$\mathfrak{su}(2n\text{-}6)$\\ 
1*&4&1&4&1& 2 \\\hline
\end{tabular} \, .
\end{equation}

\begin{table}[h!]\begin{center}
\renewcommand{\arraystretch}{1.5}
\begin{tabular}{|c|c|c|c|c|}
\hline
NU model & $\mu(a)$ & $\mu(b)$ & $\mu(c)$ & $\mu(d)$\\
\hline \hline
$[\kod{I}_{n-p-0}]$  &0&0&$n+p$&$n+p$ \\ \hline 
 $[\kod{I}_n-\kod I_p^{\ast}]$  &2&3&$6+n+p$&$6+n+p$ \\ \hline
 $[\kod{II}_{n-p}]$  &2&3&$5+n+p$&$6+n+p$ \\ \hline 
\end{tabular}
\caption{Parabolic type 4 models.}
  \label{tab:parabolic4}\end{center}\end{table}

\subsection{Parabolic type 5}

The final class in the NU list is that of parabolic type 5 models,
which includes just 6 degenerations. Only the two of them collected in table \ref{tab:parabolic5}
admit a smooth resolution presented in the following.  
The parabolic type 5 $[\kod{II}_{n-p}]$ is not the same as
the one listed in table \ref{tab:parabolic4}. Their sextics are distinct and lead to different resolutions. 
For the $E_8\times E_8$  heterotic string, the resolution of  the type 5 $[\kod{II}_{n-3}]$, $n>3$,  yields
{\footnotesize
\begin{align}
\label{resoIInp5}
&
\begin{tabular}{|ccccccccccc|}
\hline
&  $\mathfrak{su}(2)$ &$\mathfrak{so}(7)$& $\mathfrak{su}(2)$ & $\mathfrak{so}(12)$&$\mathfrak{sp}(3)$& $\mathfrak{so}(14)$ &$\mathfrak{sp}(3)$&$\mathfrak{so}(14)$& $\mathfrak{sp}(3)$\\
1 &      2 &3&1&4 &  1&4& 1& 4& 1 \\\hline
\end{tabular}\,
\begin{tabular}{|cc|}
\hline
$\mathfrak{so}(14)$& $\mathfrak{sp}(3)$\\
4& 1
\\\hline
\end{tabular}^{\, \oplus (n-4)} \hspace*{-5mm}\times \nonumber\\
& \times 
\begin{tabular}{|cccccccccc|}
\hline
$\mathfrak{so}(13)$&$\mathfrak{sp}(2)$ &$\mathfrak{so}(11)$  
  & $\mathfrak{sp}(1)$&  $\mathfrak{so}(9)$ &&$\mathfrak{g}_2$&$\mathfrak{sp}(1)$&&\\
4& 1& 4&  1 & 4&1&3&2&2&1*
\\\hline
\end{tabular} \, ,
\end{align}
}
whereas the resolution of the same model in the $\mathrm{Spin}(32)/\ZZ_2$ heterotic string gives
\begin{equation}
\label{modII43type5}
\begin{tabular}{|cccccc|}
\hline
 &  $\mathfrak{sp}(n\text{+}1)$ & &  & & \\
 & 1  & &  & & \\
  &  $\mid$ & &  & & \\
$\mathfrak{sp}(n\text{+}8)$ &  $\mathfrak{so}(4n\text{+}20)$ &$\mathfrak{sp}(2n\text{+}3)$& $\mathfrak{so}(4n\text{+}8)$&
  $\mathfrak{sp}(2n\text{-}3)$&$\mathfrak{su}(2n\text{-}2)$\\ 
1*&4&1&4&1& 2 \\\hline
\end{tabular} \, .
\end{equation}
The above results are evidently different from the resolutions of the type 4 $[\kod{II}_{n-3}]$ displayed in \eqref{resoIInp2} and  \eqref{modIIn3}.

\begin{table}[h!]\begin{center}
\renewcommand{\arraystretch}{1.5}
\begin{tabular}{|c|c|c|c|c|}
\hline
NU model & $\mu(a)$ & $\mu(b)$ & $\mu(c)$ & $\mu(d)$\\
\hline \hline
$[\kod{I}_{n-p-q}]$  &0&0&$n+p+q$&$n+p+q$ \\ \hline 
$[\kod{II}_{n-p}]$  {\scriptsize$p=2k+l$, $ l=0,1$ }&2&3&$5+l+2k+n$&$6+l+2k+n$ \\ \hline
\end{tabular}
\caption{Parabolic type 5 models.}
  \label{tab:parabolic5}\end{center}\end{table}

\section{Final comments}
In this work, we have studied heterotic compactifications with six dimensional T-fects 
leaving an $E_8\times E_7$ or a  $\mathrm{Spin}(28) \times SU(2)/\ZZ_2$ subgroup  unbroken. 
We have focused on configurations which are 
locally described by a $T^2$ fibration over a complex
one-dimensional base with a smooth, up to the degeneration points, $SU(2)$ structure bundle, patched
together using arbitrary elements of $SO^+(2,3,\mathbb{Z})$---an order
four subgroup of the T-duality group $O(2,3,\mathbb{Z})$. Generically, this gives
rise to backgrounds without a global classical geometric
interpretation. At certain points in the base, the fibration, or
bundle data on it, will degenerate and will no longer have---in any
T-duality frame---an interpretation in terms of the heterotic string
on a smooth $T^2$ with a smooth vector bundle on top. Our goal in this
paper has been to characterize the physics arising from such singular
points.

We have exploited the fact that for backgrounds preserving
$E_8\times E_7$ or $\mathrm{Spin}(28) \times SU(2)/\ZZ_2$,
the geometric data of the heterotic string
on $T^2$ can be encoded in the geometry of a genus-two (sextic)
Riemann surface. One can then define vacua 
by fibering this genus-two Riemann surface over a complex one-dimensional
base. For monodromies in $SO^+(2,3,\mathbb{Z})$, or equivalently
$Sp(4,\ZZ)$, one can classify the ways in which such fibration can
degenerate \cite{ogg66,Namikawa:1973yq}. Using heterotic/F-theory
duality to reinterpret these degenerations of the sextic as
degenerations of dual F-theory K3s fibered over the
same base, we are able to read off the low-energy physics at the degeneration
point.

We performed a systematic analysis on the full set of sextic degenerations. Remarkably, we found that many (non-)geometric degenerations
are described by the same low-energy physics. Often these are given by 
the long-understood configurations of pointlike instantons sitting on ADE singularities. 
It would be very interesting to understand the origin of this phenomenon in heterotic
language without going to F-theory.

A second notable finding is that not all of the possible
sextic degenerations admit an F-theory dual that can be smoothed
out in a crepant way by a finite number of blow-ups. 
As explained in section~\ref{sec:resolution}, this follows from the fact that in these 
cases the F-theory configuration is associated with non-minimal Weierstra\ss{}
models in complex codimension one (after we performed already some base blow-ups). 
In these cases, we cannot determine the low-energy physics using F-theory techniques, since the 
dynamics of F-theory on such backgrounds is unknown. 
Assuming that these vacua are consistent too, it would be very important to find out which kind of 
theories arise from these backgrounds in the IR.
They may give rise to free 
or trivial theories, or alternatively to interacting SCFTs without a tensor branch---at least no geometrically manifest tensor branch.
Understanding these `non-minimal models' is an open problem that deserves further attention.

\bibliography{papers}
\bibliographystyle{utphys}

\end{document}